\documentclass[reprint,showpacs,preprintnumbers,amsmath,amssymb,nofootinbib,groupedaddress,superscriptaddress,showkeys,twocolumn]{revtex4-1}

\usepackage{graphicx}
\usepackage{dcolumn}
\usepackage{epsfig}
\usepackage[dvips]{color}
\usepackage{bm}
\usepackage{fancybox}
\usepackage{mathrsfs}
\usepackage{booktabs}
\usepackage{framed}
\definecolor{shadecolor}{gray}{0.90}

\allowdisplaybreaks[4]

\begin{document}

\title{
MeV scale leptonic force for cosmic neutrino spectrum
and muon anomalous magnetic moment
}

\author{Takeshi Araki}
\email{araki@krishna.th.phy.saitama-u.ac.jp}
\affiliation{%
Department of physics, Saitama University,
\\
Shimo-Okubo 255, 
338-8570 Saitama Sakura-ku, 
Japan
}

\author{Fumihiro Kaneko}
\affiliation{%
Department of physics, Saitama University,
\\
Shimo-Okubo 255, 
338-8570 Saitama Sakura-ku, 
Japan
}

\author{Toshihiko Ota}
\email{toshi@mail.saitama-u.ac.jp}
\affiliation{%
Department of physics, Saitama University,
\\
Shimo-Okubo 255, 
338-8570 Saitama Sakura-ku, 
Japan
}

\author{Joe Sato}
\email{joe@phy.saitama-u.ac.jp}
\affiliation{%
Department of physics, Saitama University,
\\
Shimo-Okubo 255, 
338-8570 Saitama Sakura-ku, 
Japan
}

\author{Takashi Shimomura}
\email{shimomura@cc.miyazaki-u.ac.jp}
\affiliation{%
Faculty of Education and Culture, 
Graduate School of Education, 
Miyazaki University,
\\
Gakuen-Kibanadai-Nishi 1-1,
889-2192 Miyazaki,
Japan
}

\date{\today}


\preprint{\bf UME-PP-002, STUPP-15-223}


\begin{abstract}
Characteristic patterns of cosmic neutrino spectrum
reported by the {\sf IceCube} Collaboration
and long-standing inconsistency between theory and experiment
in muon anomalous magnetic moment are simultaneously 
explained by an extra leptonic force mediated by a gauge field 
with a mass of the MeV scale.
With different assumptions for redshift distribution 
of cosmic neutrino sources, 
diffuse neutrino flux is calculated
with the scattering between cosmic neutrino and cosmic neutrino
background through the new leptonic force.
Our analysis sheds light on a relation among lepton physics 
at the three different scales, PeV, MeV, and eV, 
and provides possible clues to the distribution of sources of 
cosmic neutrino and also to neutrino mass spectrum.
\end{abstract}

\maketitle

\section{Introduction}

Astrophysics and neutrino physics entered a new era after 
the discovery of high-energy cosmic neutrino events 
observed by the {\sf IceCube} Collaboration~\cite{%
Aartsen:2013bka,Aartsen:2013jdh}.
The follow-up 
reports~\cite{Aartsen:2014gkd,Aartsen:2014muf,Aartsen:2015ivb,Aartsen:2015ita}
with additional statistics uncover the spectrum of cosmic neutrino
in the energy range between $\mathcal{O}(10)$ TeV and 
$\mathcal{O}(1)$ PeV.
The reported spectrum shows some remarkable features, for example,
(i) the neutrino flux diminishes steeply as the energy increases,
and the best-fit spectral index is $s_{\nu} = 2.5$~\cite{Aartsen:2015ita};
and (ii) there is a gap in the energy range between
400 TeV and 1 PeV~\cite{Aartsen:2014gkd,Aartsen:2015ita}.
The high spectral index is consequent on the sudden end of the observed 
event spectrum at the high-energy edge ($E_{\nu}\simeq 3$ PeV) and high 
event rate at the low-energy bin 
$E_{\nu} \lesssim 100$ TeV~\cite{Aartsen:2015ivb,%
Aartsen:2014muf,Palomares-Ruiz:2015mka,Fong:2014bsa,Aartsen:2015ita,Chen:2014gxa}.
In contrast, in Refs.~\cite{Loeb:2006tw,Murase:2013rfa,Tamborra:2014xia}, 
it is also pointed out that
neutrino spectrum from standard hadronuclear process 
($pp$ inelastic scattering) with the spectral index higher than 
$s_{\nu} \gtrsim 2.2$
causes a serious conflict with the gamma-ray observation at {\sf Fermi-LAT}, 
if the spectrum does not have cutoff at the low energy.
The presence of the gap may also urge reconsideration of the assumption 
of simple power-law spectrum, which typically results from $pp$ reaction 
in the so-called cosmic-ray reservoir such as the galaxy cluster.
Although these features in the spectrum have not been conclusive
in statistics, plenty of attempts have been made to reproduce them 
from the aspects of both astrophysics~\cite{He:2013cqa,Murase:2013rfa,Liu:2013wia,Dado:2014mea,Tamborra:2014xia,Chakraborty:2015sta}
(for a review, see Ref.~\cite{Murase:2014tsa}) 
and particle
physics~\cite{Feldstein:2013kka,Esmaili:2013gha,Ibarra:2013cra,Bai:2013nga,Ema:2013nda,Bhattacharya:2014vwa,Zavala:2014dla,Higaki:2014dwa,Bhattacharya:2014yha,Ema:2014ufa,Rott:2014kfa,Esmaili:2014rma,Kopp:2015bfa,Murase:2015gea,Ahlers:2015moa,Esmaili:2015xpa,Aisati:2015vma,Roland:2015yoa,Anchordoqui:2015lqa,Berezhiani:2015fba,Boucenna:2015tra,Ko:2015nma,Ioka:2014kca,Ng:2014pca,Ibe:2014pja,Blum:2014ewa,Araki:2014ona,Cherry:2014xra,Kamada:2015era,DiFranzo:2015qea,Barger:2013pla,Dutta:2015dka}.

In Refs.~\cite{Ioka:2014kca,Ng:2014pca,Ibe:2014pja,Blum:2014ewa,Araki:2014ona,Cherry:2014xra,Kamada:2015era,DiFranzo:2015qea},
the origin of the gap in the observed spectrum was asked to the 
attenuation of cosmic neutrino, which is caused by the scattering with 
cosmic neutrino background (C$\nu$B) through a new interaction between 
neutrinos, the so-called neutrino secret
interaction~\cite{BialynickaBirula:1964zz,Bardin:1970wq,Bilenky:1999dn}.\footnote{%
The resonant neutrino-neutrino scattering mediated by the standard model 
(SM) $Z$ boson has been investigated 
in the context of the $Z$-burst scenario to address 
ultrahigh-energy cosmic neutrinos~\cite{Weiler:1982qy,Weiler:1983xx,Roulet:1992pz,Yoshida:1994ci,Yoshida:1996ie,Weiler:1997sh,Sigl:1998vz,Fodor:2001qy,Fodor:2002hy,Kalashev:2002kx,Eberle:2004ua,Barenboim:2004di}.
For early works on the relation between 
the secret interaction and new mediation fields,
see, e.g., Refs.~\cite{Kolb:1987qy,Keranen:1997gz,Goldberg:2005yw,Baker:2006gm,Hooper:2007jr}.
The effect on the cosmic neutrino spectrum from 
the scattering between neutrino and dark matter 
is discussed 
in Refs.~\cite{Cherry:2014xra,Davis:2015rza,DiFranzo:2015qea}.
}
In such a scenario, the narrow width of the gap can be
explained by the resonant behaviour of the scattering.
In the previous study~\cite{Araki:2014ona},
we introduced a new gauged leptonic force to explain the gap
and pointed out that the leptonic force could simultaneously
explain the disagreement between theory and experiment
in muon anomalous magnetic moment.
We improve in this paper our numerical method and 
calculate diffuse neutrino flux, taking account of the distribution 
of the source of cosmic neutrino with respect to the redshift.
Moreover, we search through the model parameter space to find a set of 
parameters that can reproduce not only the gap but also the sharp edge 
at the upper end of the cosmic neutrino spectrum.
The existence of the edge is expected to improve the fit of the 
spectrum with a lower value of spectral index to the observation. 
Here we also discuss constraints on the model, some of which were not 
considered in our previous study, such as the neutrino-electron 
scattering process, invisible decay of a light particle at colliders,
big bang nucleosynthesis (BBN), and supernova cooling.

The paper is organized as follows:
In the second section, we describe our model and illustrate parameter 
regions relevant to the cosmic neutrino spectrum and muon anomalous 
magnetic moment.
We also discuss constraints from laboratory experiments and cosmological 
and astrophysical observations.
Differential equations for diffuse neutrino flux with the leptonic force 
are given in Sec.~\ref{Sec:calc}.
The spectra calculated with different model parameters and redshift 
distribution of cosmic neutrino sources are compared with the 
observation in Sec.~\ref{Sec:num}. 
Finally, we discuss the relation between the characteristic features 
of the cosmic neutrino spectrum and neutrino mass spectrum, and also 
the distribution of sources of cosmic neutrino.

\section{Model and constraints}
We extend the SM of particle physics with 
a massive vector boson $Z'$ that mediates a new leptonic force,
\begin{align}
 \mathscr{L}_{\text{int}}
 =
 g_{Z'}^{}
 Q_{\alpha \beta}
 \left[ \overline{L}_{\alpha} \gamma^{\rho} L_{\beta}
 +
 \overline{\ell_{R}}_{\alpha} \gamma^{\rho} \ell_{R_\beta}
 \right] Z'_{\rho},
 \label{eq:Lint}
\end{align}
where $L_{\alpha}$ and $\ell_{R \alpha}$ are a lepton doublet and
a right-handed charged lepton singlet with flavour 
$\alpha =\{e, \mu,\tau \}$, respectively.
We choose the flavour structure of the interaction as
$Q_{\alpha \beta} = \text{diag}(0,1,-1)$, which corresponds to the 
$U(1)$ gauge interaction associated with muon number minus tau 
number ($L_{\mu} - L_{\tau}$)~\cite{Foot:1990mn,He:1990pn}.
In this paper, we do not discuss the details of the model, such as 
a mechanism of the symmetry breaking.
\footnote{%
Although we do not fully describe the model Lagrangian, 
we assume that the Yukawa sector also respects the symmetry,
and it is broken so as not to shift the mass eigenbasis 
of charged leptons, i.e., the lepton flavour is not violated.
The gauge symmetry and its phenomenology have been 
discussed in
Refs.~\cite{Foot:1994vd,Choubey:2004hn,Ota:2006xr,Heeck:2010pg,Heeck:2011wj,Harigaya:2013twa,Altmannshofer:2014cfa,Kile:2014jea,Heeck:2014qea,Crivellin:2015mga}. 
}
Instead, we handle the two parameters, 
the coupling $g_{Z'}$ and the mass $M_{Z'}$
of the gauge boson, as parameters that describe the model.

The interaction with neutrinos in Eq. (\ref{eq:Lint}) 
is expected to produce the gap and the edge in the cosmic neutrino spectrum 
through the resonant scattering with C$\nu$B.
In the resonant scattering process 
$\nu_{\text{Cosmic}} \bar{\nu}_{\text{C$\nu$B}} \rightarrow \nu \bar{\nu}$ 
mediated by $Z'$, only cosmic neutrinos having the energy corresponding
to the resonance energy $E_{\text{res}}$ are selectively scattered off
by C$\nu$B on the way from its source 
to the {\sf IceCube}~\cite{Ioka:2014kca,Ng:2014pca,Ibe:2014pja,Blum:2014ewa,Araki:2014ona,Kamada:2015era,DiFranzo:2015qea}, 
which results in the gap around $E_{\rm res}$.
Here, $E_{\text{res}}$ is given as
\begin{align}
 E_{\text{res}} = \frac{M_{Z'}^{2}}{2 m_{\nu} (1+z)}
\label{eq:Eres}
\end{align}
where $m_\nu$ stands for a mass of the target C$\nu$B and 
$z$ is the redshift parameter at which the scattering occurs.
In Eq.~\eqref{eq:Eres}, the C$\nu$B is assumed to be at rest.
With an assumption of $m_\nu = {\cal O}(0.1)$ eV, 
the scale of $M_{Z^\prime}$ can be estimated as $M_{Z^\prime} = {\cal
O}(1 - 10)$ MeV for $E_{\rm res} \simeq 1$ PeV.
Meanwhile, in order to scatter a sufficient amount of cosmic 
neutrino during the travel of $\mathcal{O}(1)$ Gpc, 
the size of the cross section is required to be larger than $\sim
10^{-30}$ cm$^{2}$ at the resonance~\cite{Ioka:2014kca,Ng:2014pca}.
In the $L_\mu - L_\tau$ model, 
the cross section near the resonance is estimated 
as\footnote{%
In the numerical calculation, 
the scattering cross section in the neutrino mass eigenbasis is used,
cf. Eqs.~\eqref{eq:diff-eq-nu} and \eqref{eq:diff-eq-antinu}.
}
\begin{align}
 \sigma_{\text{res}} 
 =  \frac{2 \pi g_{Z'}^{2} 
 }{M_{Z'}^{2}}
 \delta \left(
 1 - \frac{M_{Z'}^{2}}{s}
 \right)
\label{eq:crosssection-estimation}
\end{align}
by using a $\delta$-function approximation, 
where $s \simeq 2m_{\nu}E_{\nu}$ is the square of 
the center-of-mass energy 
in the limit of small C$\nu$B momentum.
The requirement to the cross section turns out to be 
$g_{Z'} \gtrsim \mathcal{O}(10^{-4})$.
Putting it all together, the model parameter region 
that is relevant to the cosmic neutrino spectrum 
at the energy range around $1$ PeV
can be deduced as
\begin{eqnarray}
g_{Z^\prime} \gtrsim \mathcal{O}(10^{-4})
~~{\rm and}~~
M_{Z^\prime} = \mathcal{O}(1\sim 10)~{\rm MeV}.
\label{eq:param-region-IceCube}
\end{eqnarray}

\begin{figure}[t]
\unitlength=1cm
\begin{picture}(8.5,5)
\includegraphics[width=8cm]{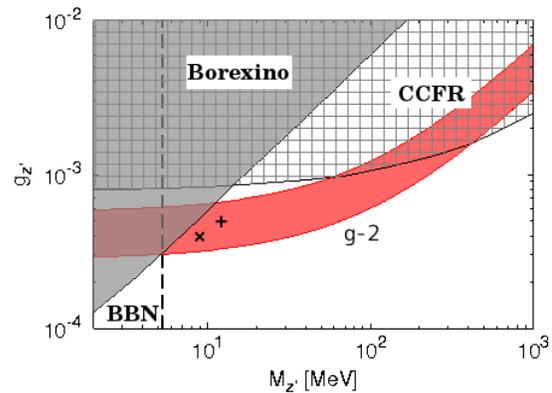}
\end{picture}
\caption{%
Relevant parameter region and the constraints.
The red band represents a parameter region favoured by muon anomalous 
magnetic moment within $2\sigma$.
The hatched region is excluded by the lepton trident search at the 
{\sf CCFR} experiment.
The region excluded by the measurement of $\nu e \rightarrow \nu e$  
at {\sf Borexino} is filled with a gray colour. 
The vertical dashed line stands for the lower bound on $M_{Z^\prime}$ 
from BBN.
Two symbols $+$ and $\times$ indicate 
$(M_{Z^\prime},~g_{Z^\prime})=(11~{\rm MeV},~5\times 10^{-4})$ 
and $(9~{\rm MeV},~4\times 10^{-4})$, respectively, which are used 
in Sec. IV. 
See the text for details.
}
\label{Fig:MZ-gZ}
\end{figure}
The interaction with charged leptons (namely muons) in 
Eq. (\ref{eq:Lint}) is responsible for an extra contribution to muon 
anomalous magnetic moment.
The measurement~\cite{Bennett:2006fi} deviates from the SM 
predictions~\cite{Davier:2010nc,Hagiwara:2011af,Aoyama:2012wk,Kurz:2014wya} 
by around 3$\sigma$.
The extra contributions from various types of new physics to fill 
this discrepancy have been intensively studied~\cite{Czarnecki:2001pv},
and also in the context of the $L_{\mu} - L_{\tau}$
model~\cite{Baek:2001kca}.
The $Z'$ contribution with the combinations of $g_{Z'}$ and $M_{Z'}$
indicated with the red band in Fig.~\ref{Fig:MZ-gZ}
successfully reproduces the observed value of muon anomalous magnetic 
moment within $2\sigma$ errors. 
In Fig.~\ref{Fig:MZ-gZ}, we also show some experimental 
constraints (see below) on the model,  
which pin down the model parameter region onto
\begin{align}
 g_{Z'} \sim \mathcal{O}(10^{-4})
 \text{ at } 
 M_{Z'} = {\cal O}(10)\text{ MeV}.
\label{eq:MZ-gZ-for-g_mu}
\end{align}
It is worth noting that this parameter region 
has some overlap with the region Eq.~\eqref{eq:param-region-IceCube} 
required 
to reproduce the gap and the edge 
in the cosmic neutrino spectrum.

In the following, we summarize experimental constraints on the 
$L_\mu - L_\tau$ model.
\vspace{0.2cm}
\\
\noindent
$\circ$~{\bf Neutrino trident production process}:
In the mass region ($M_{Z'} \lesssim 100$ MeV) we are working with, 
the tightest constraint comes from searches for the neutrino trident 
process: the production process of a $\mu^- \mu^+$ pair 
with a muon neutrino, which results from the scattering 
of a muon neutrino from a target nucleus.
The process was observed at several neutrino beam experiments, 
e.g., the {\sf CHARM-II}~\cite{Geiregat:1990gz} 
and the {\sf CCFR}~\cite{Mishra:1991bv}. 
Since the reported results are in good agreement with 
the SM prediction, which is mediated by the $Z$ and 
the $W$ bosons,
an additional contribution to the process mediated by $Z'$,
whose amplitude is proportional to $g_{Z'}^{2}$, 
is strongly constrained.
Indeed, the constraint on $M_{Z^\prime}$ and $g_{Z'}$ was recently 
evaluated in Ref.~\cite{Altmannshofer:2014pba}.
We adopt the $95\%$ C.L. bound of the {\sf CCFR} experiment,
which is shown in Fig. \ref{Fig:MZ-gZ} as the hatched region. 
\vspace{0.2cm}
\\
\noindent
$\circ$~{\bf Neutrino-electron scattering}:
Although the electron is not charged under the $L_\mu - L_\tau$ symmetry, 
$Z^\prime$ can interact with an electron through a kinetic mixing 
$\epsilon$ between $Z'$ and a photon, which is induced by loop diagrams.
The total contribution to the kinetic mixing $\epsilon$
in the $L_{\mu} - L_{\tau}$ model 
is finite, and it is estimated as
\begin{align}
 |\epsilon_{\text{loop}}| 
 = \frac{8}{3} \frac{e g_{Z'}}{(4\pi)^{2}}
 \ln \frac{m_{\tau}}{m_{\mu}}
 =
 7.2 \cdot 10^{-6}
 \left(
 \frac{g_{Z'}}{5 \cdot 10^{-4}}
 \right),
\label{eq:kinetic-epsilon}
\end{align}
where $e$ is the elementary electric charge.
This leads to an extra contribution to 
the elastic $\nu e \rightarrow \nu e$ scattering signal 
in the solar neutrino measurement at the {\sf Borexino} experiment.
The amplitude is proportional 
to $\epsilon e g_{Z'}/(q^{2} - M_{Z'}^{2})$,
where $q^{2}$ is the momentum transfer.
The constraints from {\sf Borexino} are discussed in
Refs.~\cite{Harnik:2012ni,Agarwalla:2012wf,Bilmis:2015lja} 
in the context of 
various different scenarios of new leptonic force.
Here, we interpret the bounds given in Ref.~\cite{Harnik:2012ni}
with those in the $L_\mu - L_\tau$ model 
while taking account of 
the fraction of mass eigenstates of solar neutrino, 
which is given in Ref.~\cite{Nunokawa:2006ms}.
In Fig.~\ref{Fig:MZ-gZ}, the excluded region is filled with 
a gray colour.
The measurement of $\nu$-$e$ elastic scattering at 
{\sf LSND}~\cite{Auerbach:2001wg} places a similar bound to 
the coupling $g_{Z'}$ at $M_{Z'} \simeq 10$ MeV
(and a weaker bound at $M_{Z'} \simeq 1$ MeV)~\cite{Bilmis:2015lja}. 
\vspace{0.2cm}
\\
\noindent
$\circ$~{\bf Beam dump experiments}: 
Once the kinetic mixing between a photon and $Z^\prime$ appears,
$Z^\prime$ can be produced on shell at beam dump experiments,
whose production rate is $\mathcal{O}(\epsilon^{2} e^{2})$.
However, the $Z^\prime$ immediately decays into the 
neutrino-antineutrino channel in the $L_{\mu} - L_{\tau}$ model.
Therefore, the constraints from the search for $e^+ e^-$ pairs
at the detectors located downstream of the beam dump,
such as E137~\cite{Bjorken:1988as,Bjorken:2009mm}, are 
irrelevant for our scenario.
\vspace{0.2cm}
\\
\noindent
$\circ$~{\bf Invisible decay of a light particle}:
Since $Z'$ in this framework decays dominantly to $\nu \bar{\nu}$,
one of the smoking-gun signals in collider experiments is
a light particle decaying to an invisible mode.  
The kinetic mixing leads the on-shell production of $Z'$.
The process,
$e^{+} e^{-} \rightarrow \gamma Z' \rightarrow \gamma +\text{invisible}$,
is searched at {\sf BABAR}~\cite{Aubert:2008as},
which sets the bound to the coupling between the electron and $Z'$ 
at $\sim 10^{-3}$~\cite{Essig:2013vha}.
Our reference choices of parameters satisfy this condition.
\vspace{0.2cm}
\\
\noindent
$\circ$~{\bf BBN}: 
A constraint on $M_{Z^\prime}$ is derived from BBN.
If $Z^\prime$ is as light as the temperature at the era of BBN, 
its existence increases the number of relativistic degrees 
of freedom, $N_{\rm eff}$, and the success of the standard BBN 
might be spoiled, which leads the lower bound 
$M_{Z^\prime} \gtrsim 1$ MeV \cite{Ahlgren:2013wba}. 
This condition is always satisfied on the parameter region 
of our interest.
Nevertheless, $Z^\prime$ with a mass of ${\cal O}(10)$ MeV 
may indirectly contribute to $N_{\rm eff}$
through a raise in the temperature of 
$\nu_{\mu}$ and $\nu_{\tau}$~\cite{Kamada:2015era}.
In Fig.~\ref{Fig:MZ-gZ}, we display the lower bound on $M_{Z^\prime}$ 
from the indirect contribution with $\Delta N_{\rm eff}<0.7$ as 
the vertical dashed line, 
which is taken from Ref.~\cite{Kamada:2015era}.
\vspace{0.2cm}
\\
\noindent
$\circ$~{\bf Other constraints and future improvement}:
There exist many other experimental observations constraining 
the $L_\mu - L_\tau$ model.
Most of them, however, are either irrelevant in the mass region 
we are focusing on (i.e. $M_{Z^\prime} < 100$ MeV) or weaker 
than the bounds discussed above.
Here, we mention some of them;
searches for the SM $Z$ boson decay to four leptons at the {\sf LHC} 
can be used to constrain the $L_\mu - L_\tau$ 
model~\cite{Heeck:2011wj,Harigaya:2013twa}. 
However, it becomes insensitive in the mass region below $M_{Z^\prime} \simeq 10$ 
GeV~\cite{Altmannshofer:2014pba} 
due to the invariant mass cut for the same-flavour leptons.
Furthermore, $Z^\prime$ alters the decay rates of $Z$, $W$, 
and mesons, but constraints from them are weaker than the {\sf CCFR} 
bound~\cite{Laha:2013xua,Lessa:2007up}.
The bound from a precise measurement of cosmic microwave background 
anisotropy is much weaker than those discussed above, 
cf. Fig.~1 in~\cite{Ng:2014pca} and discussion 
in Refs.~\cite{Cyr-Racine:2013jua,Archidiacono:2013dua}

It should be noted that a severe constraint on $g_{Z^\prime}$ possibly 
arises from supernova neutrino observations~\cite{Kamada:2015era}.
The existence of $Z^\prime$ with an MeV-scale mass interrupts 
free streaming of $\nu_\mu$ and $\nu_\tau$ in the core, which could 
make the diffusion time much longer than the estimation 
from the standard supernova cooling process.
To avoid possible problems, an introduction of an additional 
invisible particle, e.g. the QCD axion, might be necessary.

Apart from astrophysical/cosmological observations and
the constraints through the kinetic mixing term,
a direct test of a muonic force mediated by a boson with a sub-GeV
mass is planned at CERN SPS~\cite{Gninenko:2014pea},
which is expected to improve the sensitivity
to the coupling $g_{Z'}$ by orders of magnitude
and fully cover the parameter region referred
with Eqs.~\eqref{eq:param-region-IceCube} and \eqref{eq:MZ-gZ-for-g_mu}.
We expect that the trident events have been recorded 
at near detectors in modern long-baseline neutrino 
oscillation experiments, which may be accessible
in the present moment.
This might already give us an opportunity to explore 
the relevant parameter region.

\section{Diffuse neutrino flux}
\label{Sec:calc}
In order to calculate diffuse neutrino flux $\phi_{\nu_{i}}$
observed at {\sf IceCube},
we numerically solve the simultaneous partial differential equations 
with respect to differential number density $\tilde{n}_{\nu_{i}} (E_{\nu_{i}}, z)$ 
of cosmic neutrino $\nu_{i}$,
which are given in Refs.~\cite{Farzan:2014gza,Ng:2014pca,DiFranzo:2015qea}:
\begin{align}
 \frac{\partial \tilde{n}_{\nu_{i}}}{\partial t}
 =&
 \frac{\partial }{\partial E_{\nu_{i}}} 
 b \tilde{n}_{\nu_{i}}
 +
 \mathcal{L}_{\nu_{i}}
 -
 c 
 n_{\text{C$\nu$B}}
 \tilde{n}_{\nu_{i}}
 \sum_{j}
 \sigma (\nu_{i} \bar{\nu}_{j}^{\text{C$\nu$B}} \rightarrow \nu \bar{\nu})
 \nonumber 
 \\
 &+ c  n_{\text{C$\nu$B}}
 \sum_{j,k}
 \int_{E_{\nu_{i}}}^{\infty}
 {\rm d} E_{\nu_{k}}
 \tilde{n}_{\nu_{k}}
 \frac{{\rm d} \sigma (\nu_{k} \bar{\nu}_{j}^{\text{C$\nu$B}}
 \rightarrow \nu_{i} \bar{\nu})}
 {{\rm d} E_{\nu_{i}}}
 \nonumber
 \\
 &+ c  n_{\text{C$\nu$B}}
 \sum_{j,k}
 \int_{E_{\nu_{i}}}^{\infty}
 {\rm d} E_{\bar{\nu}_{k}}
 \tilde{n}_{\bar{\nu}_{k}}
 \frac{{\rm d} \sigma (\bar{\nu}_{k} \nu_{j}^{\text{C$\nu$B}}
 \rightarrow \nu_{i} \bar{\nu})}
 {{\rm d} E_{\nu_{i}}},
 \label{eq:diff-eq-nu}
 \\
 \frac{\partial \tilde{n}_{\bar{\nu}_{i}}}{\partial t}
 =&
 \frac{\partial }{\partial E_{\bar{\nu}_{i}}} 
 b \tilde{n}_{\bar{\nu}_{i}}
 +
 \mathcal{L}_{\bar{\nu}_{i}}
 -
 c 
 n_{\text{C$\nu$B}}
 \tilde{n}_{\bar{\nu}_{i}}
 \sum_{j}
 \sigma (\bar{\nu}_{i} \nu_{j}^{\text{C$\nu$B}} \rightarrow \nu \bar{\nu})
 \nonumber 
 \\
 &+ c  n_{\text{C$\nu$B}}
 \sum_{j,k}
 \int_{E_{\bar{\nu}_{i}}}^{\infty}
 {\rm d} E_{\bar{\nu}_{k}}
 \tilde{n}_{\bar{\nu}_{k}}
 \frac{{\rm d} \sigma (\bar{\nu}_{k} \nu_{j}^{\text{C$\nu$B}}
 \rightarrow \bar{\nu}_{i} \nu)}
 {{\rm d} E_{\bar{\nu}_{i}}}
 \nonumber
 \\
 &+ c  n_{\text{C$\nu$B}}
 \sum_{j,k}
 \int_{E_{\bar{\nu}_{i}}}^{\infty}
 {\rm d} E_{\nu_{k}}
 \tilde{n}_{\nu_{k}}
 \frac{{\rm d} \sigma (\nu_{k} \bar{\nu}_{j}^{\text{C$\nu$B}}
 \rightarrow \bar{\nu}_{i} \nu)}
 {{\rm d} E_{\bar{\nu}_{i}}},
 \label{eq:diff-eq-antinu}
\end{align}
where $i,j,k=\{ 1,2,3 \}$ are the indices for neutrino mass eigenstates.
The time $t$ is related to redshift $z$ as 
$\frac{{\rm d} z}{{\rm d} t} = -(1+z) H(z)$.
Following the discussion in Ref.~\cite{Farzan:2008eg},
we treat cosmic neutrino as an incoherent sum of mass eigenstates.
The first term on the right-hand side is responsible for 
energy loss of cosmic neutrino, owing to redshift, and
the energy-loss rate $b$ is given with $b=H(z) E_{\nu}$.
The second term represents the influx from sources of cosmic neutrino.
In this study, we assume that all sources provide the same 
spectrum of cosmic neutrino, i.e., $\mathcal{L}_{\nu_{i}} (E_{\nu_{i}},z)$
is simply parametrized as $\mathcal{L}_{\nu_{i}} (E_{\nu_{i}},z)
=\mathcal{W}(z) \mathcal{L}_{0} (E_{\nu_{i}})$
with the cosmic neutrino spectrum 
$\mathcal{L}_{0}(E_{\nu})$ from each source 
and the source distribution $W(z)$ with respect to redshift $z$.
Here, the source distribution function is assumed to be common 
for all the mass eigenstates of cosmic neutrino.
We adopt a power-law spectrum, which is
characterised by the spectral index $s_{\nu}$
and the cutoff energy $E_{\text{cut}}$: 
\begin{align}
 \mathcal{L}_{0} (E_{\nu}) 
 =
 \mathcal{Q}_{0}
 E_{\nu}^{-s_{\nu}}
 \exp
 \left[
 -\frac{E_{\nu}}{E_{\text{cut}}}
 \right],
\label{eq:L0}
\end{align}
where $\mathcal{Q}_{0}$ is the normalization of the flux,
which will be adjusted so as to fit to the observed flux.
This type of spectrum typically results from 
hadronuclear process ($pp$ inelastic scattering) 
in the cosmic-ray reservoir, 
and the values of $s_{\nu}$ and $E_{\text{cut}}$ are expected
to be determined by properties (acceleration rate, i.e., magnetic 
field and size~\cite{Hillas:1985is}) of the cosmic neutrino source.
The flavour composition of cosmic neutrino from $pp$ reaction 
is expected to be 
($\nu_{e}$, $\nu_{\mu}$, $\nu_{\tau}$;
$\bar{\nu}_{e}$, $\bar{\nu}_{\mu}$, $\bar{\nu}_{\tau}$)
=(1,2,0;1,2,0) at each source,
which leads to each mass eigenstate producing 
approximately with an equal rate, i.e.,
the normalization factor $\mathcal{Q}_{0}$ is assumed to 
be common to all the mass eigenstates in our calculations.
Although the sources of the PeV cosmic neutrinos 
have not been identified yet,
we assume the following function inspired by 
the star formation rate (SFR)~\cite{Yuksel:2008cu}
as a test distribution:
\begin{align}
 \mathcal{W}(z) = \begin{cases}
	 (1+z)^{3.4} \quad 0 \leq z < 1,\\
	 (1+z)^{-0.3} \quad 1 \leq z \leq 4.	
	\end{cases}
\label{eq:sfr}
\end{align}
The third term represents the outflows caused by the scattering process
with C$\nu$B.
Here, the cross section $\sigma$ is 
\begin{align}
 \sigma(\nu_{i} \bar{\nu}_{j}^{\text{C$\nu$B}} \rightarrow \nu
 \bar{\nu})
 = \frac{|g'_{ji}|^{2} g_{Z'}^{2}}{6\pi}
 \frac{s}{(s-M_{Z'}^{2})^{2} + M_{Z'}^{2} \Gamma_{Z'}^{2}},
 \label{eq:sigma}
\end{align}
where $\Gamma_{Z'}=g_{Z'}^{2} M_{Z'}/(12 \pi)$ 
is the total decay width of $Z'$.
The coupling 
$g'_{ij} = g_{Z'} (U^{\dagger})_{i \alpha} Q_{\alpha \beta} U_{\beta j}$ 
is the coupling in the mass eigenbasis, where 
$U$ is the lepton mixing matrix.
The number density of cosmic neutrino background is given as
$n_{\text{C$\nu$B}} = 56 (1+z)^{3}$ /cm$^{3}$ 
for each degree of freedom.
The constant $c$ appearing in the third, forth, and fifth terms 
is the light speed. 
The forth and the fifth terms
provide the influx from the final states of the scattering process,
the so-called regeneration terms.
The differential cross sections for cosmic neutrino $\nu_{i}$
are calculated to be
\begin{align}
 \frac{{\rm d} \sigma(\nu_{k} \bar{\nu}_{j}^{\text{C$\nu$B}} \rightarrow
 \nu_{i} \bar{\nu})}{{\rm d} E_{\nu_{i}}}
 =& 
 \frac{|g'_{jk}|^{2} \sum_{l} |g'_{il}|^{2}}{2\pi}
 \frac{m_{\nu_{j}} E_{\nu_{i}}^{2}}{E_{\nu_{k}}^{2}}
 \nonumber 
 \\
 &\times\frac{1}{(s-M_{Z'}^{2})^{2} + M_{Z'}^{2} \Gamma_{Z'}^{2}},
 \label{eq:diff-eq-Nr1}
 \\
 \frac{{\rm d} \sigma(\bar{\nu}_{k} \nu_{j}^{\text{C$\nu$B}} \rightarrow
 \nu_{i} \bar{\nu})}{{\rm d} E_{\nu_{i}}}
 =&
 \frac{|g'_{kj}|^{2} \sum_{l} |g'_{il}|^{2}}{2\pi}
 \frac{m_{\nu_{j}} (E_{\nu_{k}} - E_{\nu_{i}})^{2}}{E_{\nu_{k}}^{2}}
 \nonumber 
 \\
 &\times\frac{1}{(s-M_{Z'}^{2})^{2} + M_{Z'}^{2} \Gamma_{Z'}^{2}}.
 \label{eq:diff-eq-Nr2}
\end{align}
For cosmic antineutrino $\bar{\nu}_{i}$, 
the differential cross section for 
the scattering process 
$\bar{\nu}_{k} \nu_{j}^{\text{C$\nu$B}} \rightarrow \bar{\nu}_{i} \nu$
($\nu_{k} \bar{\nu}_{j}^{\text{C$\nu$B}} \rightarrow \bar{\nu}_{i} \nu$)
is the same as Eq.~\eqref{eq:diff-eq-Nr1} 
(Eq.~\eqref{eq:diff-eq-Nr2}).
We numerically solve these simultaneous partial differential equations
Eqs.~\eqref{eq:diff-eq-nu} and \eqref{eq:diff-eq-antinu} 
of cosmic neutrino propagation, following the algorithms introduced 
in Ref.~\cite{TextbookFluid}. 
We confirmed that our numerical method correctly reproduces 
the results given in Ref.~\cite{Ng:2014pca}.
After the simultaneous equations are solved,
the differential number density $\tilde{n}_{\nu_{i}}$ of cosmic neutrino 
at the Earth ($z=0$) is obtained,
and 
the neutrino flux $\phi_{\nu_{i}}$ observed at {\sf IceCube} 
is calculated as
\begin{align}
 \phi_{\nu_{i}} (E_{\nu_{i}})
 =
 \frac{c}{4\pi} \tilde{n}_{\nu_{i}} (E_{\nu_{i}}, z=0).
\end{align}
In the next section, we plot the total fluxes $\Phi = \sum_{i}
(\phi_{\nu_{i}}+ \phi_{\bar{\nu}_{i}})$ as functions 
of the observed energy of cosmic neutrino.

\begin{figure*}[t]
 \unitlength=1cm
 \begin{picture}(16.5,5.5)
  \put(0,0){\includegraphics[width=5cm]{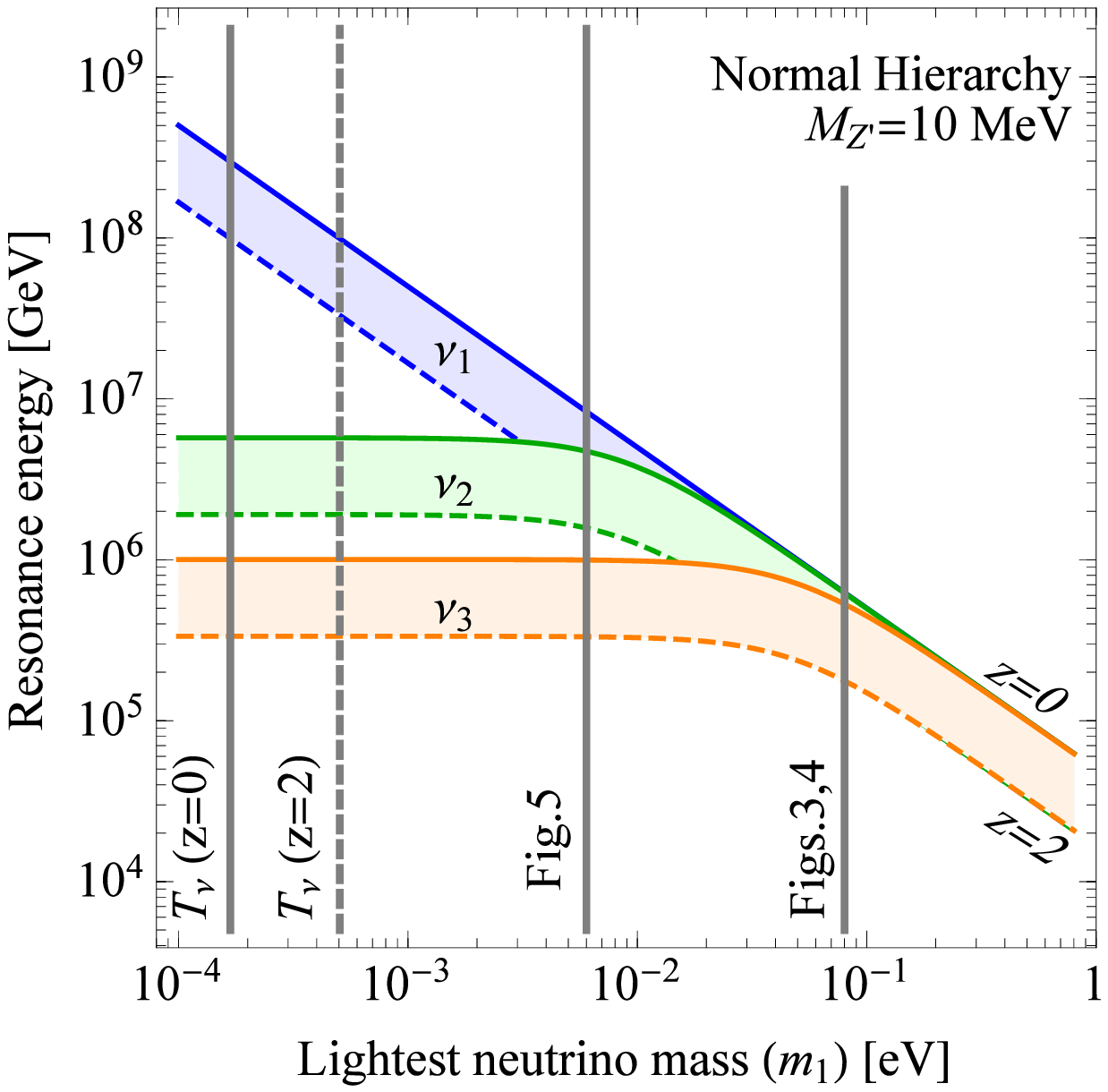}}
  \put(5.5,0){\includegraphics[width=5cm]{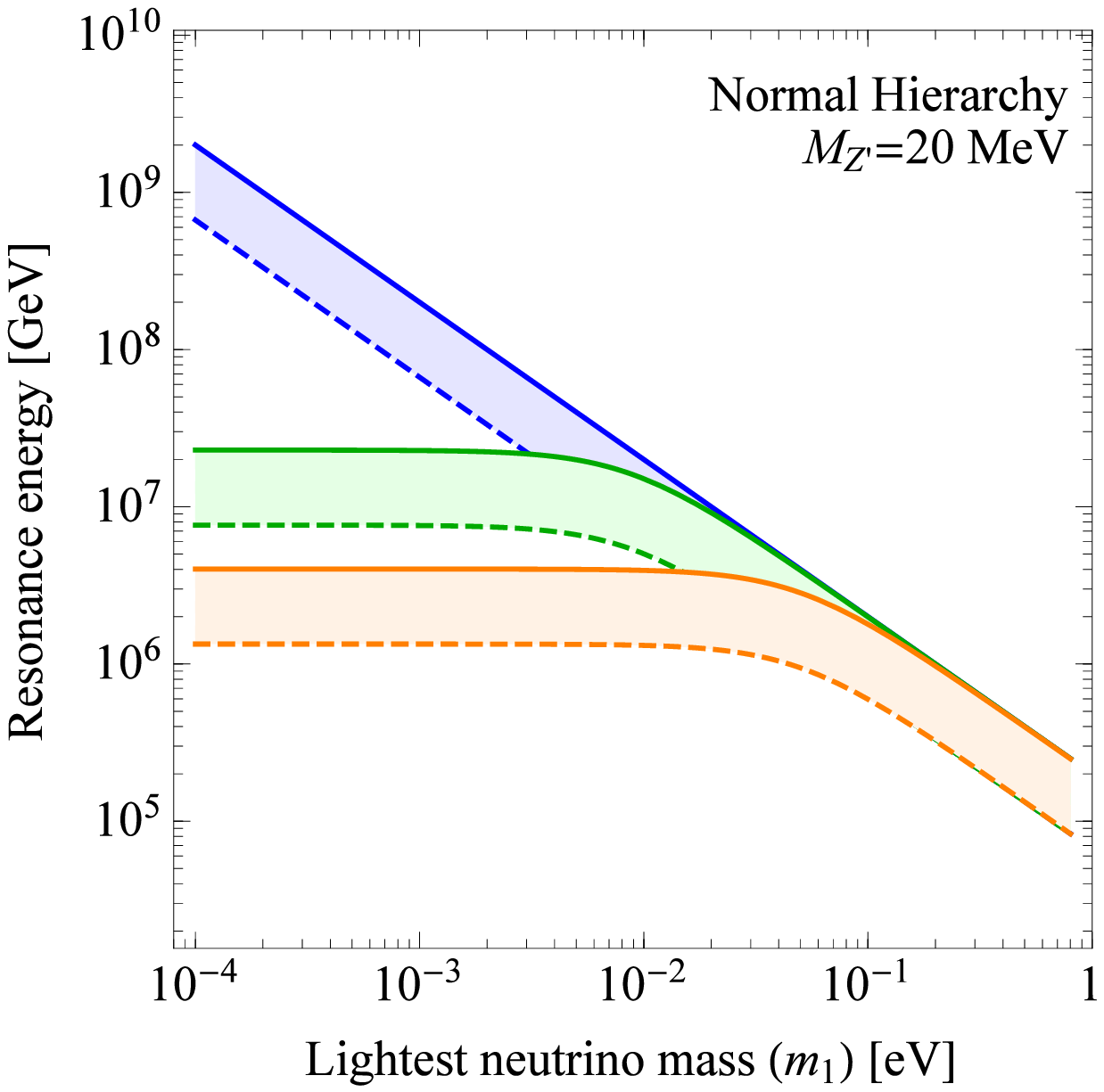}}
  \put(11,0){\includegraphics[width=5cm]{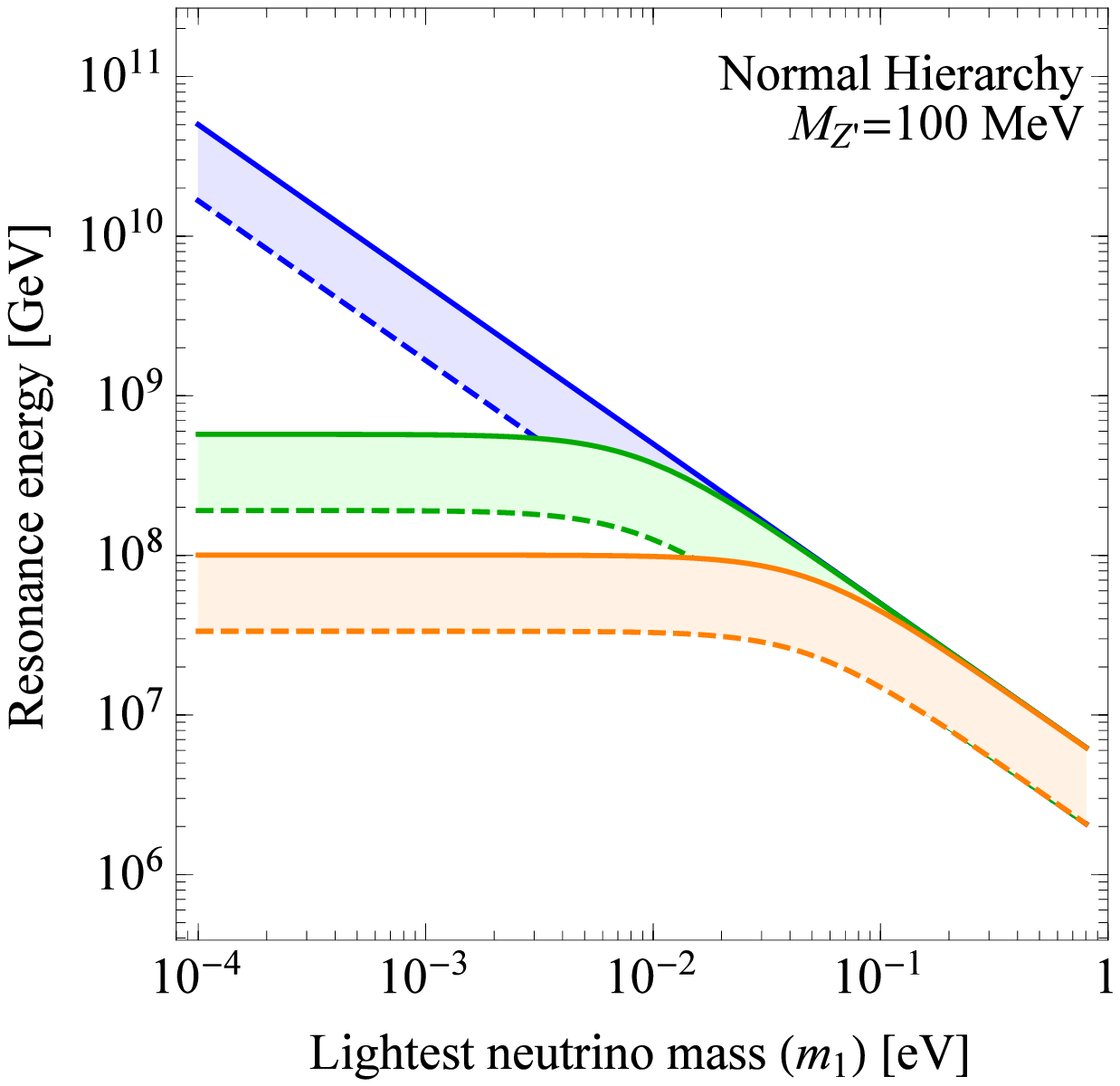}}
 \end{picture}
 \\
  \begin{picture}(16.5,5.5)
  \put(0,0){\includegraphics[width=5cm]{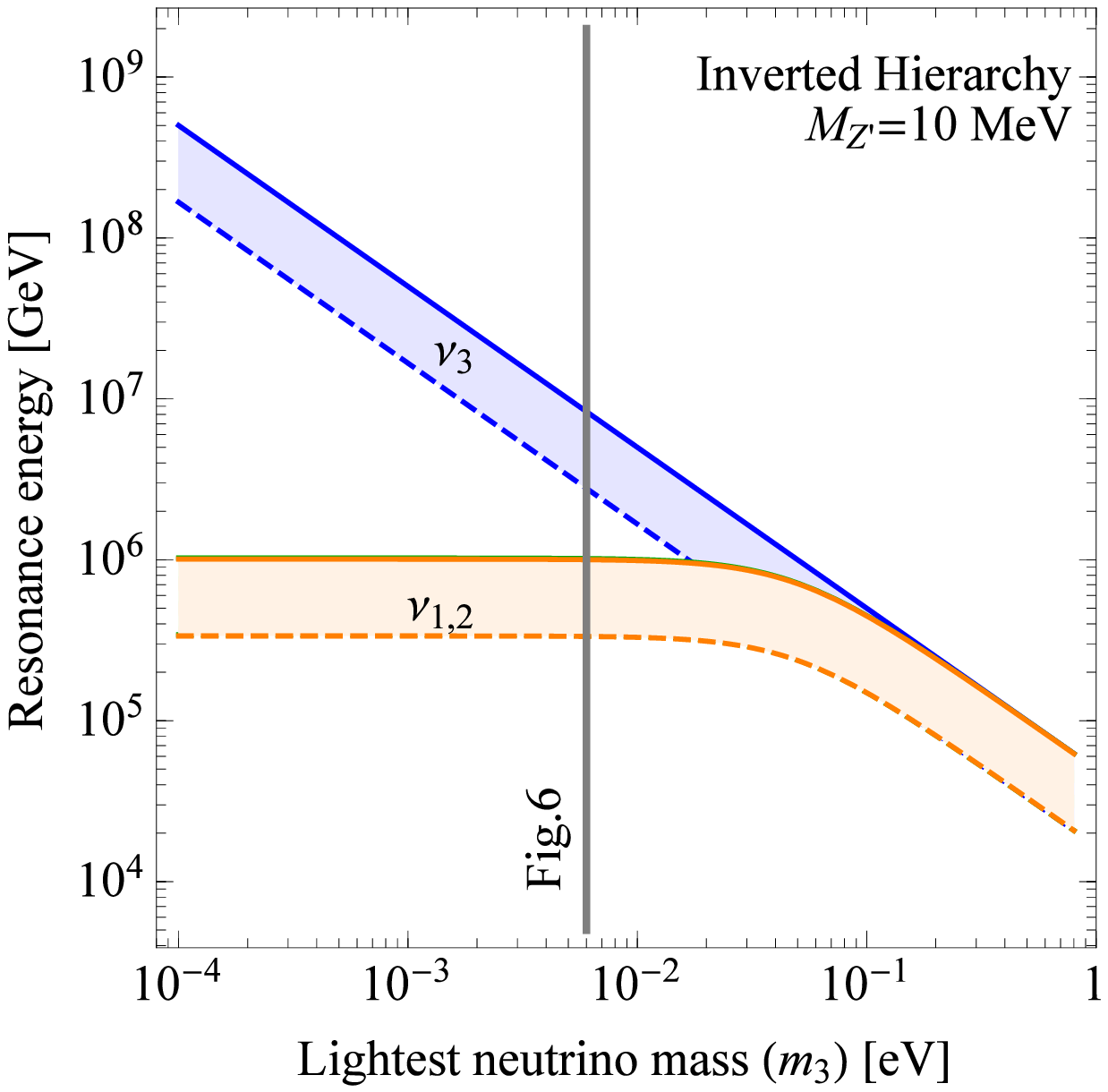}}
   \put(5.5,0){\includegraphics[width=5cm]{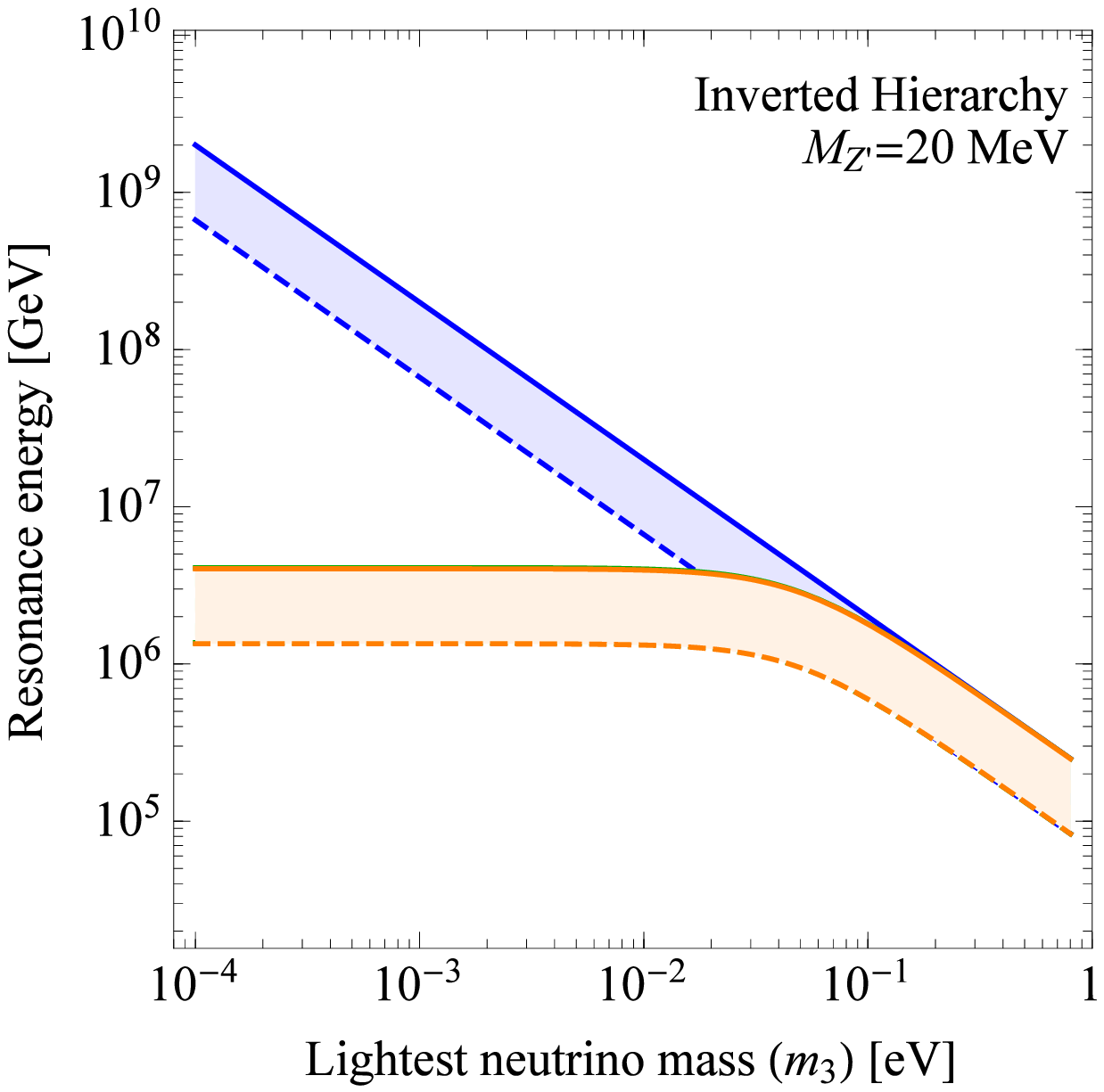}}
   \put(11,0){\includegraphics[width=5cm]{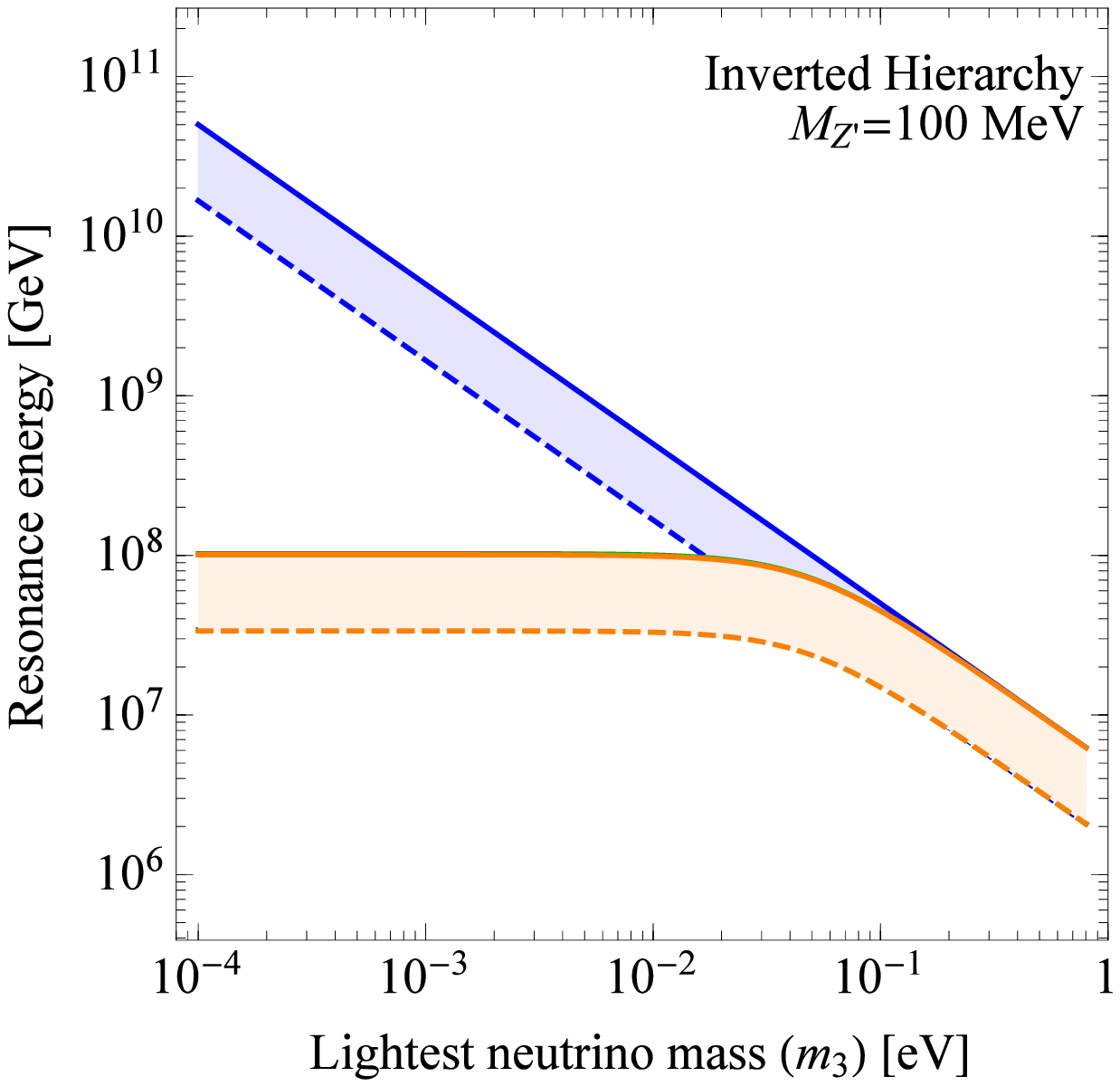}}
 \end{picture}
\caption{Resonance energies as functions of the lightest neutrino mass.
The mass $M_{Z'}$ of the leptonic gauge boson is taken to 
be $\{10,20,100\}$ MeV, and the redshift $z$ of the scattering point
is varied between $z=0$ (solid) and $z=2$ (dashed).
The band labelled with ``$\nu_{i}$'' ($i=\{1,2,3\}$) 
is the region of the resonance energy corresponding to 
the scattering with the mass eigenstate $\nu_{i}$ 
of cosmic neutrino background.
The vertical lines with $T_{\nu} (z=\{0,2\})$ indicate
the temperatures of C$\nu$B at $z=\{0,2\}$.
All the neutrino masses in our reference choices,
which are also marked with the vertical lines, 
are significantly larger than the temperatures;
The inclusion of the thermal distribution effect of C$\nu$B momentum
does not change our numerical results.
}
\label{Fig:Eres}
\end{figure*}
\section{Numerical results}
\label{Sec:num}
We numerically solve Eqs. (\ref{eq:diff-eq-nu}) and 
(\ref{eq:diff-eq-antinu}) and calculate the cosmic neutrino 
flux in the presence of the $L_\mu - L_\tau$ interaction.
Throughout the numerical study, we do not take account of
the effect of thermal distribution of the C$\nu$B momentum.
This treatment is justified by our choices of the parameters; 
we are mainly interested in the parameter region where
the lightest neutrino mass is much larger than the C$\nu$B
temperature, cf. Fig.~\ref{Fig:Eres}.
We have checked numerically that the C$\nu$B momentum effect 
does not drastically change the conclusions drawn in this section.
We come back to this point at the end of Sec. IV-A.

In the following calculations, we use the best-fit values of 
the mixing angles and the mass squared differences from 
Ref.~\cite{Forero:2014bxa}:
\begin{eqnarray}
&&
\sin^2\theta_{13} = 0.0234~(0.0240),~~
\sin^2\theta_{23} = 0.567~(0.573),
\nonumber\\
&&
\sin^2\theta_{12} = 0.323,~~
\Delta m_{21}^2 = 7.60\times 10^{-5}~~[{\rm eV^2}],
\nonumber\\
&&
|\Delta m_{31}^{2}| = 2.48~(2.38)\times 10^{-3}~~[{\rm eV^2}]
\end{eqnarray}
for the normal (inverted) mass hierarchy, 
and the $CP$ violating Dirac phase is set to zero.

\subsection{Reproduction of the gap with the SFR}
\begin{figure}[h]
\unitlength=1cm
\begin{picture}(8,5.5)
\includegraphics[width=8cm]{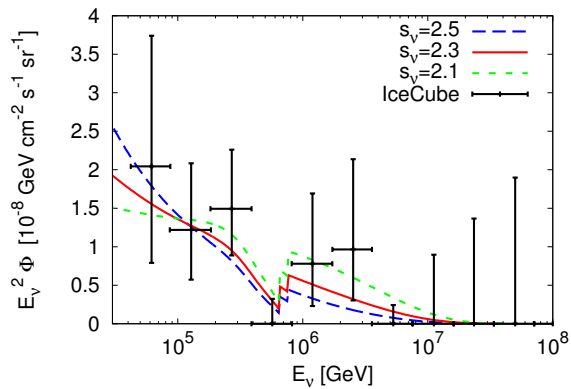}
\end{picture}
\caption{%
The cosmic neutrino fluxes 
calculated with 
the $L_{\mu} - L_{\tau}$ gauge interaction
are compared with the three-year {\sf IceCube} 
data~\cite{Aartsen:2014gkd}.
The model parameters are taken as
$M_{Z^\prime}=11$ MeV and
$g_{Z^\prime} = 5 \times 10^{-4}$.
The lightest neutrino mass 
is set to be $m_1 = 0.08$ eV 
and the normal mass hierarchy is chosen.
The SFR is assumed as the redshift distribution 
of the cosmic neutrino sources.
The cutoff energy of the original flux is placed at 
$E_{\text{cut}} = 10^{7}$ GeV.
The three different values of the spectral index $s_{\nu}$ 
are examined.}
\label{Fig:gap}
\end{figure}
To begin with, we check if the gap 
can be reproduced by 
the resonant $L_\mu - L_\tau$ scatterings.
Since neutrino consists of three generations, 
we generally expect three gaps in the spectrum.
In Fig.~\ref{Fig:Eres}, the resonance energies $E_{\text{res}}$ are
plotted as functions of the lightest neutrino mass $m_{\rm lightest}$
with different values of $M_{Z'}$ and with both the mass hierarchies.
It can be read off from the plots that 
in order to reproduce a single gap, 
the masses of neutrinos should be quasidegenerate 
so that the three resonance energies 
are located at the same point.
In Fig.~\ref{Fig:gap}, 
we show the cosmic neutrino flux with the attenuation effect
by the $L_{\mu} - L_{\tau}$ force 
and compare the results with three different values 
of the spectral index $s_\nu$.
Here we take the normal hierarchy with the lightest neutrino mass 
$m_1 = 0.08$ eV\footnote{
This leads to $\sum m_{\nu} \simeq 0.25$ eV, 
which is slightly higher than the $95\%$ C.L. from the combined 
analysis of cosmological observations~\cite{Ade:2013zuv}. 
However, once the cosmological model is extended to include more parameters, 
the constraint is expected to be relaxed.
For instance, simultaneous inclusion of $N_{\rm eff}$ and $\sum m_\nu$ 
leads to $\sum m_\nu < 0.28$ eV~\cite{Ade:2013zuv}.
}
and set the model parameters as $M_{Z^\prime}=11$ MeV 
and $g_{Z^\prime} = 5 \times 10^{-4}$.
For the sources of cosmic neutrinos, we assume the SFR, which is given 
in Eq.~(\ref{eq:sfr}), as their redshift distribution, and the cutoff 
energy $E_{\text{cut}}$, which appears in Eq. (\ref{eq:L0}), is taken 
as $E_{\text{cut}} = 10^7$ GeV. 
The normalization factor ${\cal Q}_0$ is adjusted so that 
the magnitude of the calculated flux fits the observation.
As can be seen from the figure, the flux is significantly attenuated 
around $400$ TeV $-$ $1$ PeV.
With a spectrum including the gap, one can expect
a relatively good fit to the observation,
although the gap will be shallower than the bottom of 
the calculated spectra once
the curves are averaged over each energy bin.
Since the spectrum calculated with the inverted hierarchy is 
essentially the same as the normal hierarchy shown 
at Fig.~\ref{Fig:gap}, we do not repeat it.

\begin{figure}[t]
\unitlength=1cm
\begin{picture}(8,5.5)
\includegraphics[width=8cm]{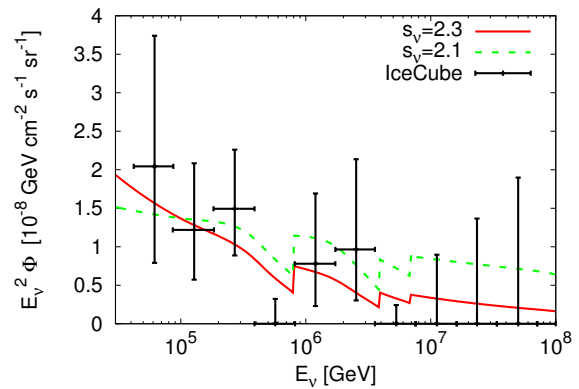}
\end{picture}
\caption{%
The cosmic neutrino flux calculated 
with $M_{Z^\prime}=9$ MeV and 
$g_{Z^\prime} = 4 \times 10^{-4}$.
Here the normal hierarchy is chosen and 
the lightest neutrino mass is set to be $m_1 = 6 \times 10^{-3}$ eV.
The spectral index is taken to be $s_{\nu} =2.3$ and $2.1$. 
}
\label{Fig:edge-NH}
\end{figure}

\begin{figure}[t]
\unitlength=1cm
\begin{picture}(8,5.5)
\includegraphics[width=8cm]{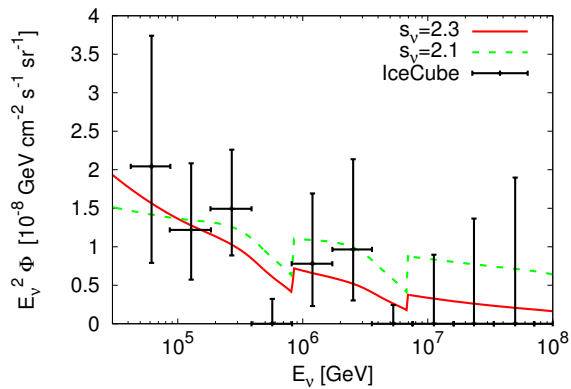}
\end{picture}
\caption{%
The same as Fig.~\ref{Fig:edge-NH} 
with the inverted mass hierarchy.
}
\label{Fig:edge-IH}
\end{figure}

Let us mention the possibility of simultaneous reproduction 
of the gap and the edge.
In view of Refs.~\cite{Loeb:2006tw,Murase:2013rfa,Tamborra:2014xia}, 
we here take lower values of $s_{\nu}$ and try to form 
the edge at the upper end of the spectrum by means of the 
$L_\mu - L_\tau$ interaction, instead of setting the cutoff 
energy by hand.
Note that with an appropriate adjustment 
of the flux normalization, lower values of the spectral index can 
still give a good fit to the current observed
spectrum~\cite{Palomares-Ruiz:2015mka,Fong:2014bsa,Chen:2014gxa}.
According to Fig.~\ref{Fig:Eres},
the mass of the lightest neutrino should be 
smaller than $10^{-2}$ eV
to split the resonance energies and distribute them 
to the positions of the gap and the edge.
The mass of $Z'$ should be smaller than $M_{Z'} \lesssim 20$ MeV 
to place the resonance energies at the appropriate positions,
cf. Eq.~\eqref{eq:Eres}.
In Fig.~\ref{Fig:edge-NH} (\ref{Fig:edge-IH}), 
we set the mass of $Z'$ to $9$ MeV,
the coupling $g_{Z^\prime}$ to $4\times 10^{-4}$,
and the lightest neutrino mass $m_1$ ($m_3$) to 
$6\times 10^{-3}$ eV with the normal (inverted) hierarchy 
of neutrino mass. 
Here, the cutoff energy is taken to be sufficiently high so 
that the numerical results do not depend on the value.
The gap is successfully reproduced 
by the scattering with the heaviest mass eigenestate of 
C$\nu$B.
On the other hand, the resonant scattering for the 
edge seems insufficient:
the flux is attenuated only between $3$ and $7$ PeV, 
which may be too narrow (and also too shallow) 
to explain the required property of the edge, 
although it is consistent with the current data.

Lastly, we comment on the effect of the C$\nu$B momentum.
If the lightest neutrino mass is chosen to be as light as 
the C$\nu$B temperature, the C$\nu$B momentum effect is 
expected to become appreciable, which would make the width 
of the edge wider.
We will study this possibility in the near future.

\subsection{Source distributions}
\begin{figure}[t]
\unitlength=1cm
\begin{picture}(8,5.5)
\includegraphics[width=8cm]{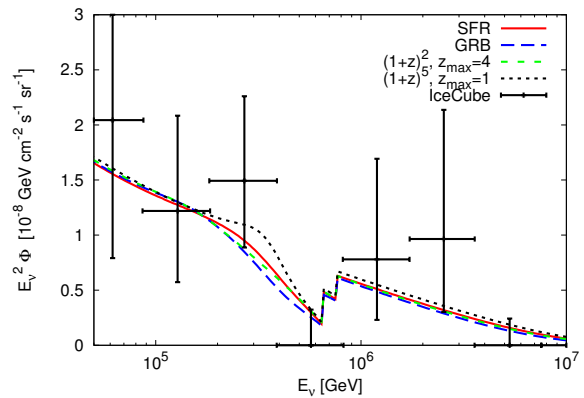}
\end{picture}
\caption{%
The same as Fig.~\ref{Fig:gap} with different types 
of source distributions. 
The spectral index is taken as $s_{\nu} =2.3$.
}
\label{Fig:dist}
\end{figure}

So far, we have adopted the SFR as the redshift distribution 
of cosmic neutrino sources in our calculations.
However, the source has not been specified yet,
and some of the astrophysical objects have been discussed 
as the candidate~\cite{He:2013cqa,Murase:2013rfa,Liu:2013wia,Dado:2014mea,Tamborra:2014xia,Chakraborty:2015sta}. 
In Fig.~{\ref{Fig:dist}}, we examine the distribution 
of gamma-ray bursts (GRBs)~\cite{Yuksel:2006qb},
\begin{eqnarray}
 \mathcal{W}_{\rm GRB}(z) \propto \begin{cases}
	 (1+z)^{4.8} \quad 0 \leq z < 1,\\
	 (1+z)^{1.4} \quad 1 \leq z \leq 4.5	
	\end{cases}
\end{eqnarray}
and the monotonic distribution~\cite{Yoshida:2012gf},
\begin{eqnarray}
 \mathcal{W}_{\rm mono}(z) \propto (1+z)^{m}
\end{eqnarray}
with $m=2,~z_{\rm max}=4$ and $m=5,~z_{\rm max}=1$, 
and compare them to the result calculated with the SFR.
Here, the spectral index is set to be $s_\nu = 2.3$.
The neutrino mass spectrum and the model parameters 
are taken to be the same as Fig.~\ref{Fig:gap}.
It is natural to conclude that,
if the power-law spectrum Eq.~\eqref{eq:L0} is assumed,
the types of source distribution do not make a big impact 
on the shape of the flux.
One can expect to find the difference caused by the choice of 
source distribution at the energy region slightly below 
the resonance point, at which the effect of regeneration 
is relevant.

\section{Discussion and conclusions}
We have introduced an anomaly-free leptonic force 
mediated by the gauge boson with a mass of the MeV scale
in order to simultaneously explain  
the two phenomena with different energy scales
in lepton physics:
(i) the disagreement between 
experimental measurement and theoretical predictions
in muon anomalous magnetic moment,
and 
(ii) 
the characteristic features of 
the cosmic neutrino spectrum reported by the {\sf IceCube}
Collaboration.
Assuming that the PeV cosmic neutrinos
are produced after the $pp$ inelastic scattering process 
in cosmic-ray reservoirs, 
we have calculated diffuse neutrino flux 
with the new leptonic force.

We have discussed the relevant constraints, such as 
the lepton trident process and the observation of 
a solar neutrino event at the {\sf Borexino}, 
and scanned the model parameter space. 
We have found the choices of parameters, which successfully 
reproduce the measured value of muon anomalous magnetic moment
and the gap between 400 TeV and 1 PeV
in the {\sf IceCube} spectrum.

Setting the mass of the leptonic gauge boson
to be around 10 MeV 
and the lightest neutrino mass to be heavier 
than $1 \times 10^{-3}$ eV,
we could arrange the three resonant energies $E_{\text{res}}$ 
corresponding to three mass eigenstates of cosmic neutrino background 
to the energy ranges of the gap ($E_{\nu} = 400$ TeV-1 PeV) 
and the edge ($E_{\nu} \simeq 3$ PeV) simultaneously.
However, the resonance at the energy corresponding to 
the edge might be too narrow (and too shallow) to explain
the sharp upper end of the spectrum, 
which is expected from the observation of the {\sf IceCube}.
If one considers the parameter region where the lightest 
neutrino mass is much lighter than $\mathcal{O}(10^{-3})$ eV,
momentum distribution of C$\nu$B begins to have an impact,
and one can expect that the inclusion of the effect
would make the resonant region wider.
We are now preparing for our next analysis in which 
we examine the thermal effect of C$\nu$B with 
smaller neutrino masses.

We have also examined the different redshift distribution of
cosmic neutrino sources and compared the resulting spectra
with each other, assuming the power-law spectrum for 
the original flux.
One can expect the visible difference in the spectrum
just below the energy region of the gap.

Before closing this paper, we make some comments on
Refs.~\cite{Kamada:2015era} and \cite{DiFranzo:2015qea} in which the
authors also considered the $L_\mu - L_\tau$ model and discussed its
impact on the cosmic neutrino spectrum as well as muon anomalous
magnetic moment.
In Ref.~\cite{Kamada:2015era}, the authors assumed coherent propagation
for cosmic neutrino and calculated the survival rate of 
neutrino flavour states for the resonant scattering 
by considering the evolution of the density matrix.
In contrast, we assume that parent particles, i.e., cosmic rays, are disturbed by the environment and do not decay into ``free space'' at the source of cosmic neutrino, which leads us to the equation of incoherent (mass eigenstate) propagation.
With the cross sections Eqs.~(\ref{eq:sigma})-(\ref{eq:diff-eq-Nr2}) 
in mass eigenbasis, we calculate the
diffuse neutrino flux by solving 
the neutrino propagation equations 
Eqs.~(\ref{eq:diff-eq-nu}) and (\ref{eq:diff-eq-antinu}).
Namely, our treatment of cosmic neutrino and approach are different from
those adopted in Ref.~\cite{Kamada:2015era}.
Both the treatments of neutrino propagation reproduced
the gap in the cosmic neutrino spectrum.
However, the flavour ratios resulting from the two methods 
are distinct from each other:
More electron neutrino can survive with the coherent propagation.
In addition, we investigate the dependence of the spectral index 
and the source distribution on the flux.
%
In Ref.~\cite{DiFranzo:2015qea}, the same propagation equations 
are used, and the diffuse neutrino flux is calculated for several 
values of neutrino masses and a $Z^\prime$ mass.
Among their parameter choices, the authors paid special attention 
to the case where the scattering effect becomes 
significant.
To illustrate it, 
they set $M_{Z'} \simeq 1$ MeV and $m_{\text{lightest}} \lesssim 10^{-3}$
eV, where the effect of C$\nu$B temperature made the resonant region 
wide.
On the other hand, we choose $M_{Z'} \simeq 10$ MeV 
and $m_{\text{lightest}} \gtrsim 10^{-3}$ eV, 
which is tailor made for the reproduction 
of the gap and the edge.
With this parameter choice, 
where the C$\nu$B temperature effect 
is not essential at all, 
we have further investigated the dependence of 
the spectral index and the source distribution 
on the flux.
In this sense, our study is complementary 
with Ref.~\cite{DiFranzo:2015qea}.
Furthermore, we have discussed the experimental 
bounds on the leptonic force 
and included the constraints from the loop-induced 
$\nu$-$e$ scattering process~\cite{Harnik:2012ni} and 
the BBN~\cite{Kamada:2015era}, 
both of which disfavour the possibility of the $Z'$ being 
as light as $\mathcal{O}(1)$ MeV.
We have confirmed that 
our numerical method correctly reproduced 
the results of Ref.~\cite{DiFranzo:2015qea}
in the parameter region with $m_{\text{lightest}} \gtrsim
10^{-2}$ eV.

\begin{acknowledgments}
We gratefully acknowledge insightful comments of 
Prof. Shigeru Yoshida.
We thank Prof. Masahiro Ibe and Dr. Ayuki Kamada
for valuable comments on constraints to the model,
Prof. Kazunori Kohri for pointing out the importance 
of the relation between cosmic neutrino and 
gamma-ray observations,
and 
Dr. Irene Tamborra for helpful comments 
on cosmic neutrino propagation.
T.O. is grateful to the Mainz Institute for Theoretical Physics (MITP)
for its hospitality and support and expresses a special 
thanks to the organisers and participants of 
{\it Crossroads of neutrino physics} 
for many constructive comments,
particularly, 
Prof. Andr\'{e} de Gouv\^{e}a and Prof. Joachim Kopp 
for useful comments on the constraints to the leptonic interactions,
Prof. Zurab Berezhiani and Dr. Claudia Hagedorn 
for discussion on the symmetry and its breaking,
and 
Prof. Pasquale Serpico for his helpful and encouraging comments 
and suggestions.
This work is supported by JSPS KAKENHI Grants 
No.~26105503 (T.O.),
No.~24340044, No.~25105009 (J.S.), 
and 
No.~15K17654 (T.S.).
\end{acknowledgments} 

\bibliography{./IceCubeGap}
\bibliographystyle{apsrev}

\end{document}